\documentclass[aps,pre,onecolumn,groupedaddresss,showpacs]{revtex4-1}		% reprint=2 
\usepackage{amsmath,amssymb}
\usepackage{color}
\usepackage{epsfig}
\usepackage{graphicx}	% Include figure files
\usepackage{subfig}
\usepackage{soul}
\usepackage{ulem}
\usepackage{hyperref}	% add hypertext capabilities
\begin{document}

%\preprint{}

\title{Complexity Study of a Single Particle Under q-Deformed Potentials} % 
%Force line 
%breaks 
%with 
%\\



%\author{F. Nutku}
%\email{fnutku@istanbul.edu.tr}
%\author{E. Ayd{\i}ner}
%\email{ekrem.aydiner@istanbul.edu.tr}

\author{Ferhat Nutku$^{1}$, K. D. Sen$^{1,2}$ and Ekrem 
Aydiner$^{1}$}\email{ekrem.aydiner@istanbul.edu.tr}

\affiliation{$^{1}$Department of Physics, Faculty	of Science, 
\.{I}stanbul University, Vezneciler, \.{I}stanbul, 34134, Turkey\\ 
$^{2}$School of Chemistry, University of Hyderabad, Hyderabad 500 046, India}

%\thanks{}


\date{\today}% It is always \today, today,
\begin{abstract}
We have studied the variation of the position space statistical complexity
measure defined by L\'{o}pez-Ruiz, Mancini, and Calbet such 			
as the product of exponential of the Shannon information entropy and the 
disequilibrium by using the 1D-normalized 
probability densities derived from solutions of the Schr\"{o}dinger equation
corresponding to the $q$-deformed harmonic oscillator and q-deformed Morse
potentials. An analysis of the numerical results in terms of Shannon 
information entropy, disequilibrium and complexity measure are presented. In 
q-deformed harmonic oscillator, q-dependence of the complexity
shows a minimum point for all excited energy levels.
In the case of q-deformed Morse Potential, 
complexity decreases with increasing $q$ for the investigated diatomic 
molecules.
\end{abstract}

% PACS, the Physics and Astronomy Classification 
%Scheme.

\maketitle

\section{Introduction}\label{sec:intro}

Complexity measures are being increasingly employed in order to understand the behavior 
of systems encountered in several disciplines of scientific inquiry. Everybody knows that 
it is very difficult to define a universal definition of complexity for all systems. 
Therefore, it can be seen that different measures for complexity have been proposed 
in the literature \cite{Lloyd2001}. A few examples of them are 
(i) algorithmic complexity \cite{Chaitin1966,Kolmogorov1968}, 
(ii) a measure of the self-organization capacity of a system \cite{Georgiev2015},
(iii) Crutchfield and Young's complexity \cite{Crutchfield1989} etc.
For a finite many-particle system, complexity may be regarded as a measure of
it's internal order/disorder which can be represented by the corresponding information 
entropy and the distance from equilibrium which is called as disequilibrium.
In the context of electronic structure of atoms and molecules, 
a suitable measure which is so called as $C_{LMC}$ has been proposed by  
L\'{o}pez-Ruiz, Mancini and Calbet (LMC) \cite{LMC,LopezRuiz2005} 
to analysis the statistical complexity. 
This measure has been widely employed to many problems in the literature
\cite{COM,LopezRuiz2005,Sanudo2011}.
Indeed, $C_{LMC}$ allocates a multiplicative role to the 
measure of distance from equilibrium in conjunction with the information entropy
to define a measure for the complexity of a finite system.

In most of the previous studies focused on quantum systems,
the wave function of a particle has been used to obtain quantities 
such as information entropy, disequilibrium and complexity.
It is known that a Gaussian type potential leads to a Gaussian wave function,
which strongly satisfies the Heisenberg inequality.
However a non-Gaussian type potential leads to a more non-exponential 
wave function which does not strongly satisfy Heisenberg inequality.
In this case, instead of Heisenberg 
inequality, a different one such as Bialynicki-Birula and Mycielski (BBM) 
inequality \cite{Biaynicki-Birula1975} can be often preferred.
BBM inequality is a theoretic-informatic inequality which can be 
applied to many different problems.
In the complexity discussion of quantum systems, the form of the wave function is 
very important, which may present different aspects of the complex behavior 
of these systems.
One of the source of the non-exponential form is 
q-deformation over the wave function, which can emerge from
non-linear interactions and non-Markovian memorial effects within the systems 
or from the interactions between the system and its surrounding environment. 
True understanding of the effect of deformed potential can shed light on 
the deep physics of complexity of several real systems.
q-deformed potentials have been considered in physics to discuss many problems.
For example; the q-deformed hyperbolic potentials 
have been proposed firstly by Arai
\cite{Arai1991,Arai2001} and they have found several applications in various 
fields of physics and chemistry. They are used for modeling and describing 
electronic conductance in disordered metals and doped semiconductors 
\cite{Alavi2004}, phonon spectrum in $^4$He \cite{Monteiro1996},
oscillatory-rotational spectra of diatomic \cite{johal1998} and multi-atomic
molecules \cite{bonatsos1997}. Furthermore, q-deformation of the Morse potential has been 
investigated by \cite{Cooper1995,Ikhdair2009,Dobrogowska2013}.
Some recent works are the investigation and comparison of the energy spectrum of 
q-oscillator and Morse-like anharmonic potential in Ref.\,\cite{Vinh2015} 
and derivation of the exact normalized wave functions for the q-deformed screened Coulomb 
Hulthen potential in Ref.\,\cite{Hall} etc.

The aim of this work is to relate complexity of a a quantum system 
with its energy level $n$ and potential deformation parameter $q$.
Therefore, in this letter we consider two simple q-deformed potentials 
and analyze complexity of the system depending on these potentials.
The outline of article is like the following: a definition and 
meaning of L\'{o}pez-Ruiz, Mancini and Calbet complexity measure
is presented in Sec.\;2. Afterwards in Sec.\;3, $C_{LMC}$ is applied to the 
probability distribution of a single particle under 
q-deformed harmonic oscillator and q-deformed Morse potentials.
As a summary, significant results and remarks are presented in Sec.\;4.
	
\section{Complexity Measure of L\'{o}pez-Ruiz, Mancini and Calbet}

The LMC complexity measure is defined for continuous systems as the following
\cite{LopezRuiz2005},
\begin{equation}\label{eq:C}
C_{LMC} = e^{S_x}D_x
\end{equation}
where $S_{x}$ and $D_{x}$ are Shanon information entropy and disequilibrium, 
respectively. These are defined in one dimension as
\begin{equation}\label{eq:S}
S_{x}  =  -\int_{-\infty}^{\infty}\rho(x)\ln \rho(x) \ \mathrm{d}x
\end{equation}
\begin{equation}\label{eq:D}
D_{x} =  \int_{-\infty}^{\infty} \rho^2(x) \ \mathrm{d}x
\end{equation}
where $\rho(x)$ is the probability density given in terms of the wavefunction as $\rho(x) 
= |\psi(x)|^2$.
Above expressions are obtained by taking the continuous limit of the 
following expressions which are expressed in terms of a discrete 
probability distribution.
\begin{equation}\label{eq:Sd}
S_{x} = -k_B\sum_{i=1}^{N}p_i(x)\ln p_i(x)
\end{equation}
\begin{equation}\label{eq:Dd}
D_{x} = \sum_{i=1}^{N} \left(p_i-\frac{1}{N}\right)^2
\end{equation}
where $p_i$ is the probability of occupying the state $i$, and $N$
is the total number of accessible states in position space. 
For a reminder, this definition of complexity is valid only for position space.
However, a complexity measure for momentum space can be also obtained by 
applying a Fourier transformation to probability distribution 
expressed in the position space.

On the other hand, according to the LMC definition one can deduce that complexity
is like predictability $P$. LMC can be used to predict whether a system is fully 
ordered or not by using $P=1-C_{LMC}$.
For instance, 
at the statistical limit case  $C_{LMC}$ becomes zero for both totally ordered
(perfect crystal) and totally disordered (random gas) systems.
Ideal gas is totally random and one can predict its randomness.
On the other hand, perfect crystal is also totally predictable, 
because it can be constructed by a unit cell and a symmetry operation 
transformation.

It is generally assumed that $C_{LMC}$ in Eq.\,\ref{eq:C} is a well known statistical 
measure of complexity for ergodic systems. On the other hand, 
we show in this study 
that $C_{LMC}$ is also a novel candidate measure to identify complexity for q-deformed   
quantum systems. Therefore, we discuss the complexity in the whole paper caused from 
q-deformed harmonic and Morse potentials by using Eqs.\,\ref{eq:C}-\ref{eq:D}.

\section{Complexity study of q-deformed potentials}
	
\subsection{q-Deformed harmonic oscillator}\label{}
	
Now, we consider, as a first example, q-deformed quantum harmonic 
oscillator. The q-oscillator is described by the Hamiltonian
\begin{equation}
		H=\frac{\hbar\omega}{2}(a^{-}a^{+}+a^{+}a^{-})
\end{equation}
where $\omega$ is the oscillator frequency.
In the complexity study of q-oscillator $\hbar$ is set to 1.
$a^+$ and $a^{-}$ are creation and 
annihilation operators satisfying the commutation relation
\begin{equation}
	[a^{-},a^{+}]_{q}=a^{-}a^{+}-qa^{+}a^{-}=1,
\end{equation}
with the deformation parameter $q$ taking values in the interval $(0, 1)$.
The effect of annihilation and creation operators to the states are given 
as below,
\begin{eqnarray}
	a^{-}|n\rangle=\sqrt{[n]}|n-1\rangle, \;
	a^{+}|n\rangle=\sqrt{[n+1]}|n+1\rangle, \;
\end{eqnarray}
where $[n]_q=\frac{1-q^n}{1-q}$ which satisfies the limits,
$\lim_{q\rightarrow 1}[n]_q=n$
and 
$\lim_{q\rightarrow 0}[n]_q=1$.

The wave function of q-oscillator is more complex than that of simple 
quantum oscillator. However, if the calculation steps in Ref.~\cite{Eremin2006} are 
repeated, the normalized wave functions can be obtained as 
\begin{eqnarray}\label{eq:psi}
|n\rangle=\Psi_n(x)=\frac{\exp({-\frac{x^2}{2}+\frac{3}{2}{i\alpha}x})}
{\pi^{\frac{1}{4}}i^n{(1-\exp({-2\alpha^2}))}^{\frac{n}{2}}\sqrt{[n]_q!}}
\sum_{k=0}^n\frac{{(-1)}^k[n]_q!}{[k]_q![n-k]_q!}\exp({2{i\alpha}(n-k)x-k\alpha^2})
\end{eqnarray}
where $\alpha=i\sqrt{(\log q)/2}$. In the above formula $[n]_q!$ is the q-factorial
and defined as 
\begin{equation}\label{eq:qfac}
[n]_q!= \prod _{{k=1}}^{n}[k]_{q}={\frac {1-q^{n}}{1-q}}
{\frac {1-q^{{n-1}}}{1-q}}\cdots {\frac {1-q^{2}}{1-q}}{\frac {1-q}{1-q}}=
{\frac {(q;q)_{n}}{(1-q)^{n}}}.
\end{equation}
where $(q;q)_{n}$ is the q-Pochhammer symbol. On the other hand, energy eigenvalues for 
q-oscillator is given by 
\cite{Eremin2006},
\begin{equation}\label{eq:EnqHO2}
	E_n=\omega\left([n]_{q}+\frac{q^n}{2}\right)=\omega\left(\frac{[n]_{q}+[n+1]_{q} 
	}{2}\right).\;
\end{equation}
We note that, in the limit of $q \rightarrow 1$, energy eigenvalues reduce 
to  
$E_n=\omega\left(n+\frac{1}{2}\right)$ of the ordinary quantum harmonic oscillator. 
However, under a small perturbation from unity $(q = 1 - \varepsilon)$, 
the energy spectrum becomes quadratic and energy eigenvalues can be approximated as
the following, 
\begin{equation}\label{eq:EnqHO}
	E_n=\omega\left(n+\frac{1}{2}-\frac{n^2}{2}\varepsilon+O(\varepsilon^2)\right).
\end{equation}
For $\epsilon = 0$, Eq.\,\ref{eq:EnqHO} reduces to eigenvalue of the 
harmonic oscillator. We note that the q-oscillator has a nonlinear spectrum and it 
satisfies the Heisenberg uncertainty as \cite{Eremin2006}
\begin{equation}
\Delta x \Delta p=\dfrac{E_n}{\omega}=\left(\frac{2-(1+q)q^n}{2(1-q)}\right)
\leq  \left(n+\dfrac{1}{2}\right)
\end{equation}
for all states where $q$ changes in the interval of $(0,1)$.
Therefore, uncertainty of q-deformed quantum harmonic oscillator is less
than the uncertainty of quantum harmonic oscillator.
		
\begin{figure*}[!ht]
\centering
\subfloat[$n=1$]
{\includegraphics*[width=0.4\textwidth]{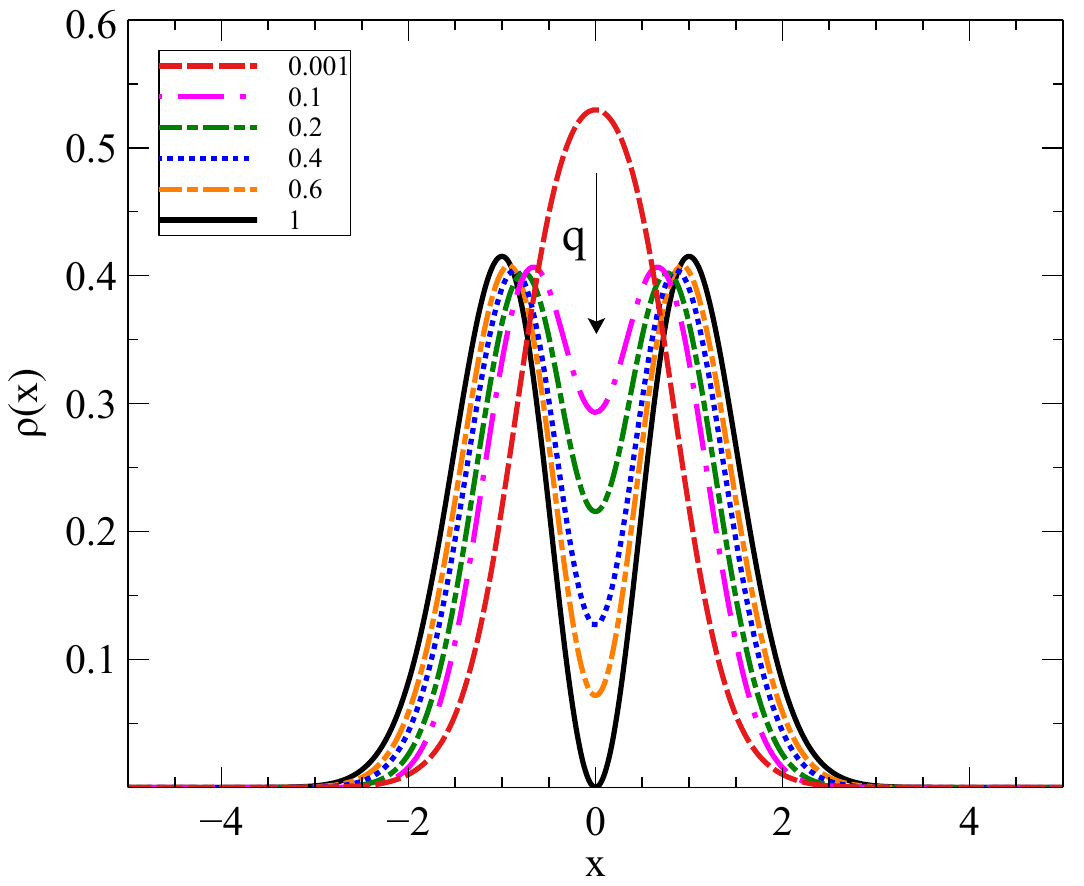}}
\subfloat[$n=2$]
{\includegraphics*[width=0.4\textwidth]{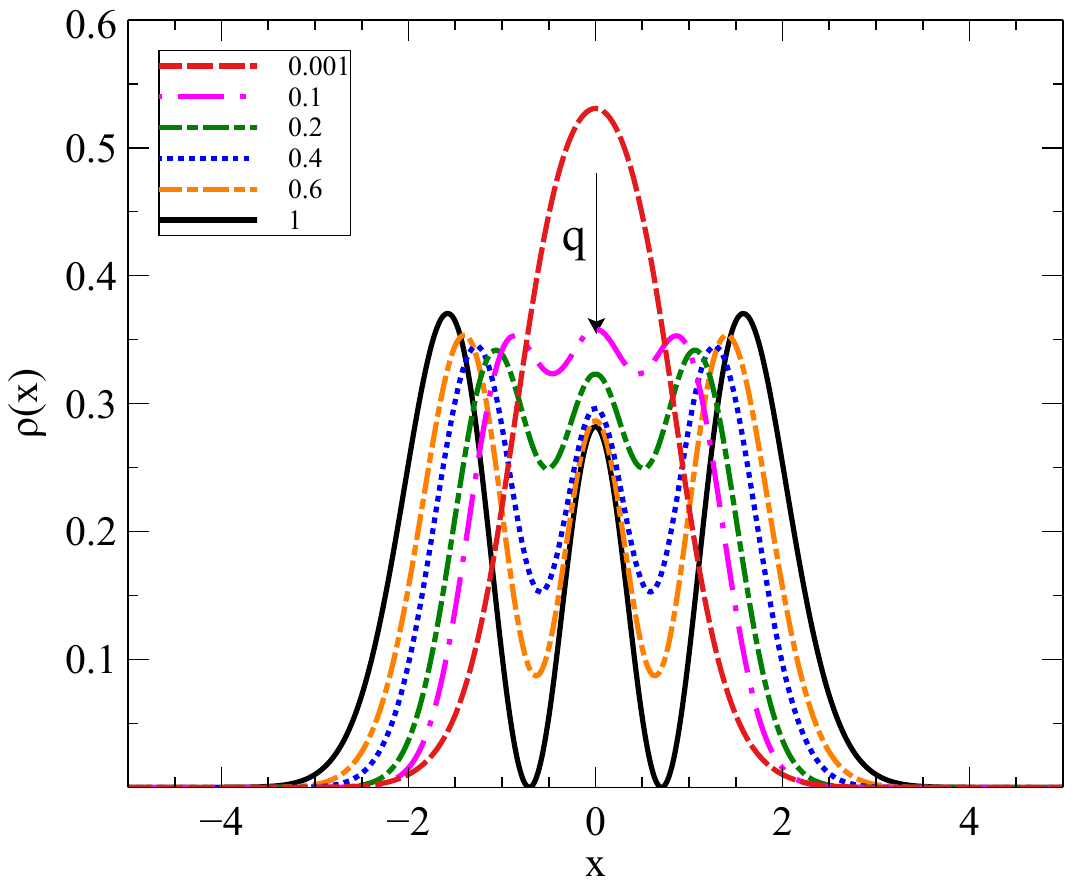}}\\
\subfloat[$n=5$]
{\includegraphics*[width=0.4\textwidth]{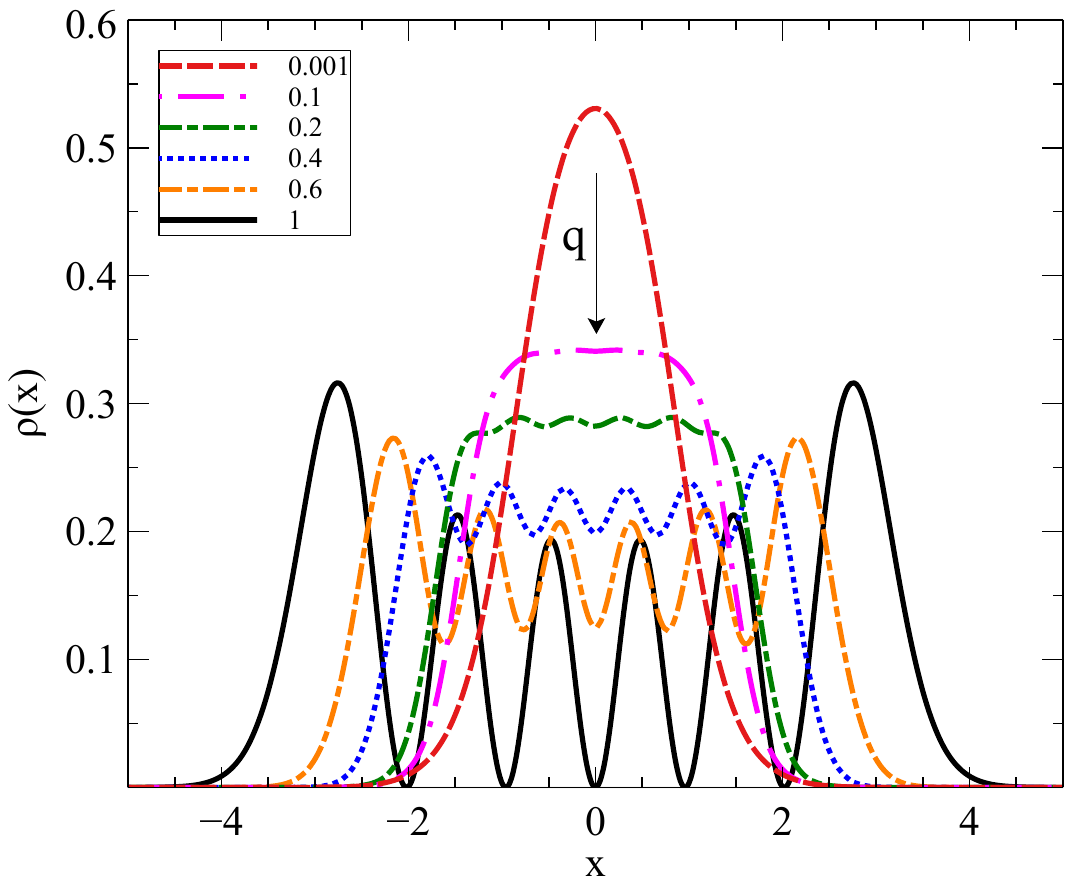}}
\subfloat[$n=6$]
{\includegraphics*[width=0.4\textwidth]{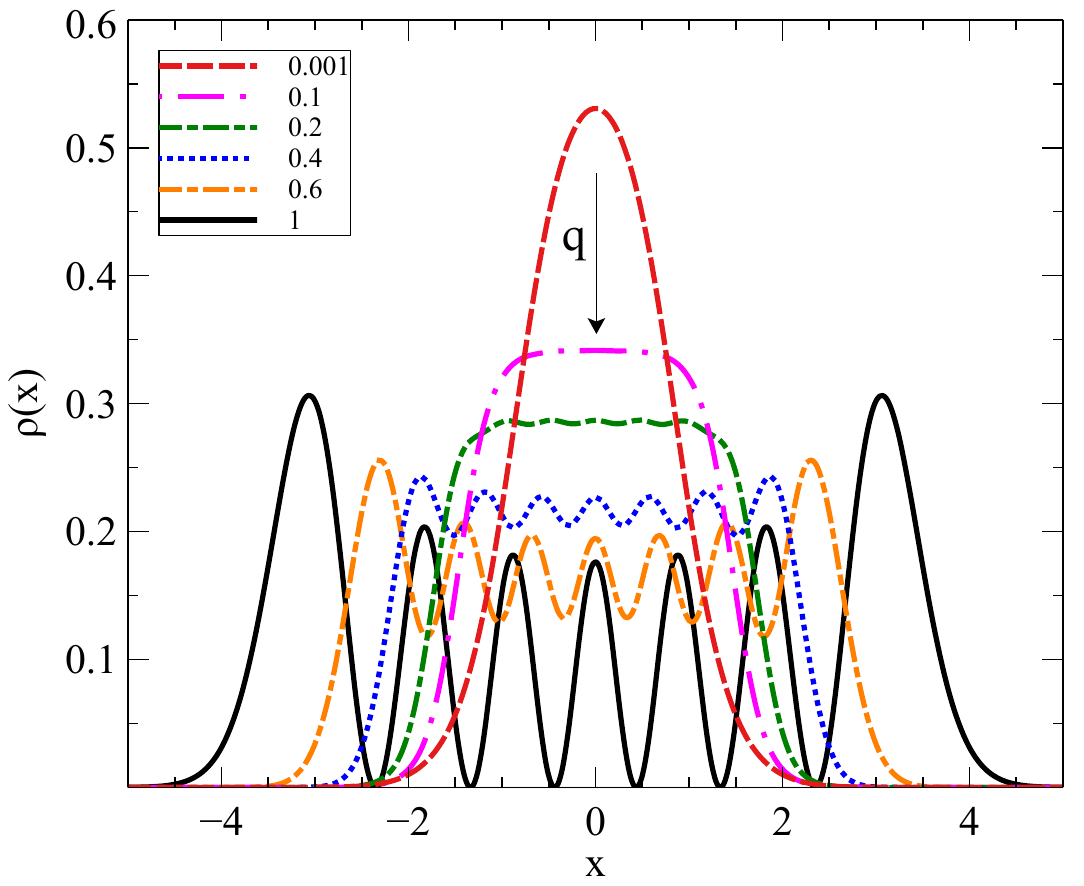}}
\caption{$q$ dependence of the probability distribution under q-deformed 
harmonic oscillator for levels $n=1,2,5,6$. The arrow indicates
increment direction of $q$.}
\label{fig:qHO_rho}
\end{figure*}
	
Applying a simple numerical procedure to the related equations, with help 
of the Eq.\,\ref{eq:psi}, probability distribution $\rho(x)$ of q-oscillator for
different energy levels can be computed. Here we present the results for $n=1,2,5,6$ to 
quantify discussing. Related numerical results for these excited states for different q 
values are given in Figs.~\ref{fig:qHO_rho}a-d. As it can be seen from 
figures, q-oscillator show very interesting behavior 
depend on both of exited states and q-parameter. The shape of the probability 
distribution dependence upon $n$ is generally known which corresponds to 
solution of the quantum harmonic oscillator i.e, q=1. Here we note that the 
number of peaks in the 
probability distribution increases with $n$. In the case of 
$q=1$, the most probable value of position for the 
lower states is very different from the classical harmonic oscillator where it spends 
more time near the boundaries of its motion. But as the quantum number increases, the 
probability distribution becomes more like that of the classical oscillator - this 
tendency to approach the classical behavior for high quantum numbers is called the 
correspondence principle.
This is the fist important point to understand of the solution of 
the quantum harmonic oscillator. We will remind this point in below discussing.
On the other hand, in the numerical study, we found that the q-dependence of 
q-oscillator is very remarkable. For $q=1$ the solution gives the standard $\rho(x)$ of 
the quantum harmonic oscillator. This is a known case as mentioned above.
However, while $q$ parameter
decreases from 1 to 0 which indicates that deformation increases
and $\rho(x)$ becomes more localized in both figures. This behavior is the 
second important point.

If we turn back to figures, we can see from  Figs.~\ref{fig:qHO_rho}a and 
\ref{fig:qHO_rho}b that 
the shape of the probability function has similar characteristic form for some q 
values. However, for very small q values this behavior of the $\rho(x)$ dramatically 
changes. For instance; at q=0.001, probability surprisingly becomes localized at the x=0. 
We say that q dominates the oscillator from quantum mechanical behavior to 
the classical Gaussian distribution one.
For q=1, there is no memory effects in the system which behaves completely quantum 
mechanical. However, when q decreases, indicating that memory effects in the system
emerges.
We can note that source of the
memory effect is the correlations between peaks of the wave functions which
might be arisen from the broken symmetry of the harmonic oscillator potential
with increasing $q$ deformation.
	
It is possible to see similar behavior in the case of higher excited states 
for example $n=5$ and 6.
The oscillating form of the $\rho(x)$ clearly changes depending on q 
values. When $q$ decreases from 1 to 0.1, 
the form gradually becomes more narrow and the oscillating 
behavior of the $\rho(x)$ between two edges decreases and vanishes so that oscillating 
form evolves into a straight oscillating plateau.
In the case of $n=1,2,5,6$, for a small $q=0.001$ value, 
as it can be seen from Fig.~\ref{fig:qHO_rho}a-d localization behavior in the the 
probability appears i.e., probability distribution becomes more Gaussian.
We note that $q$ dependence behavior of the $\rho(x)$ is a very interesting result which
probably emerges from deformation of the external potential.
These results can be concluded that when $q$ 
approaches zero, probability of the corresponding wave function of the quantum 
q-oscillator becomes localized at the origin which might be a result of
memorial or non-Markovian effects.

\begin{figure*}[h!]
		\centering
\subfloat[]
{\includegraphics*[width=0.32\textwidth]{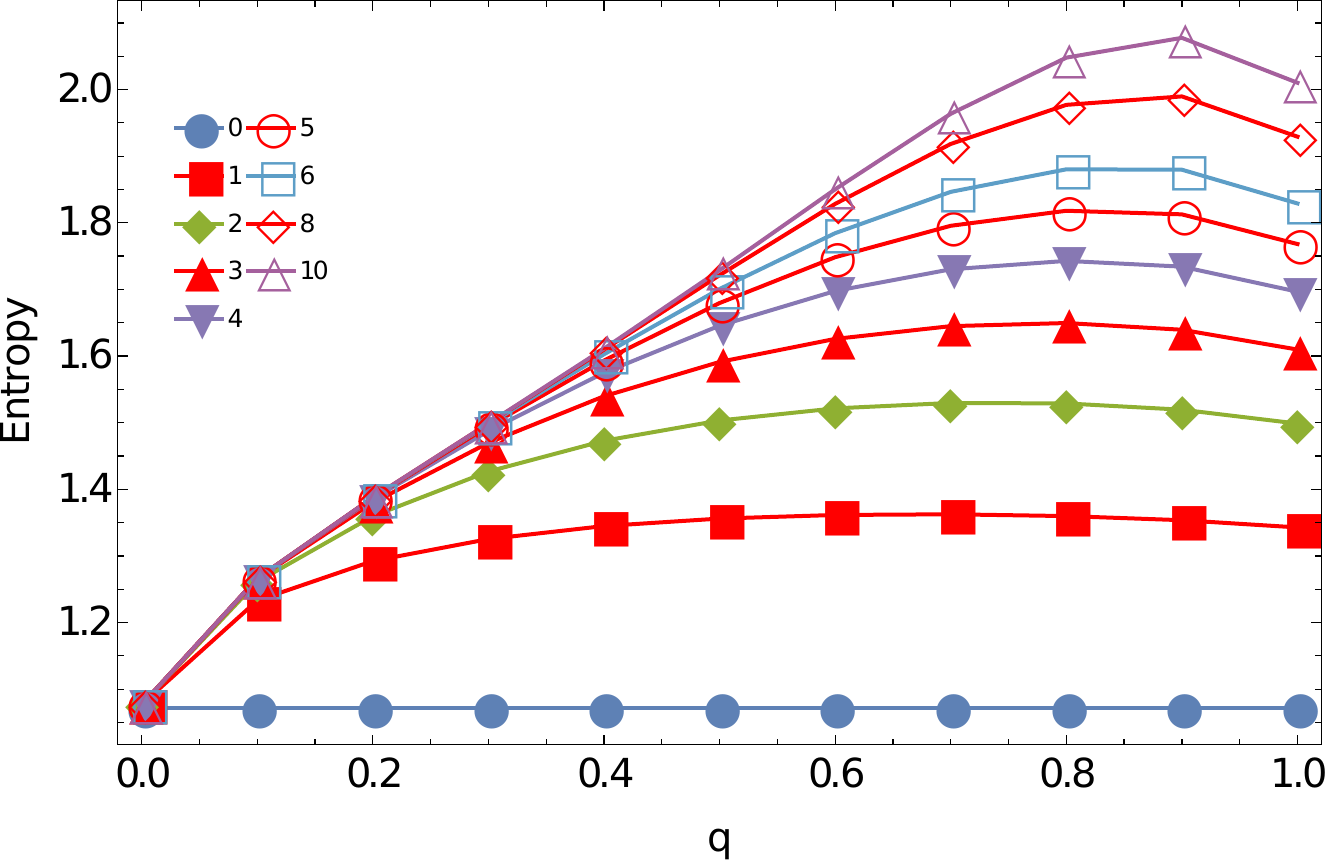}}
\subfloat[]
{\includegraphics*[width=0.32\textwidth]{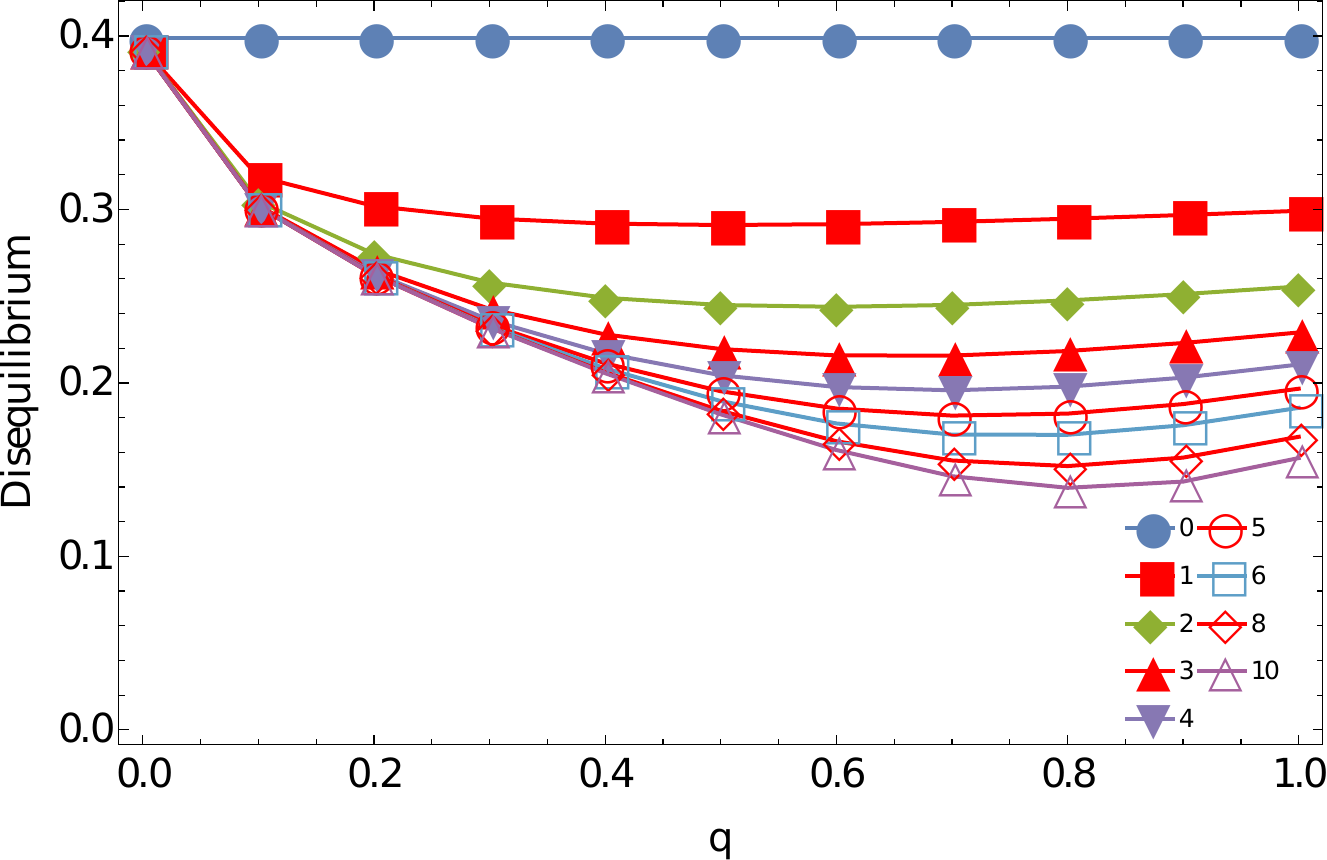}}
\subfloat[]
{\includegraphics*[width=0.32\textwidth]{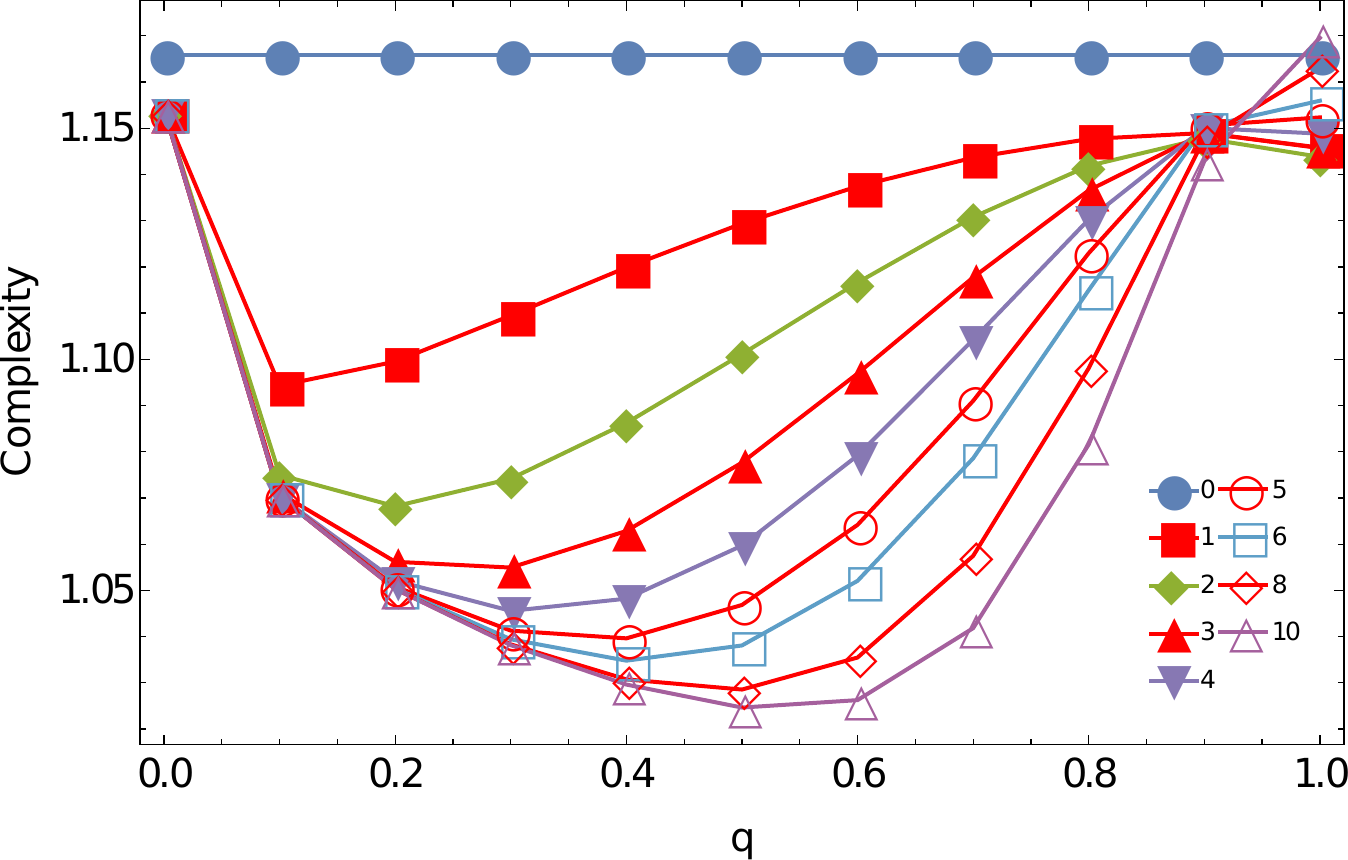}}%
\caption{$q$ dependence of the Shanon information entropy (a), 
disequilibrium (b) and complexity (c) for q-deformed harmonic
oscillator for energy levels changing in [0,10] where n=7,9 have been omitted
for preventing messiness.}
\label{fig:qHO}
\end{figure*}

So far, we have discussed the probability distribution and eigenenergies of q-oscillator.
Now we can analyze disequilibrium, information entropy and complexity properties of 
q-oscillator.
Based on Eqs.\,\ref{eq:C}-\ref{eq:D}, these quantities 
can be numerically obtained for ground and several exited states depending on q. All 
numerical results are given in  Fig.~\ref{fig:qHO}. As it can be seen from 
Fig.~\ref{fig:qHO}a that the ground state information entropy is constant 
and independent from $q$.
However, it has different shapes for all excited states. For example the 
information entropy increases and smoothly reach up to a saturation for small $n$ values 
while q increase from 0 to 1. However, this behavior rapidly changes for higher n values. 
Information entropy increases while q increases and after a critical saturation it 
decreases again between $q=0.8-1.0$. This behavior appears depending on n values.
Therefore, we see a smooth transition from quantum to classical 
behavior in information entropy.
This is a novel result. This important result shows that $q$ plays an important 
role as a memory or non-Markovian effects on the system, which drives the system from 
quantum to classical one. 
 
In Fig.~\ref{fig:qHO}b $q$ dependence of the disequilibrium is given for 
different excited states. As it can be seen from Fig.\,\ref{fig:qHO}b  disequilibrium of 
ground state is independent from q and disequilibrium has a large value.
However, disequilibrium of q-oscillator decreases with increasing q and n values, 
which means that the system passes from quantum to classical behavior with increasing q 
and n
values. On the other hand, after attaining a certain minimum value in the region
$q=0.7-0.9$, the disequilibrium is found to increase as $q$ decreases below 0.8. For a 
given value of $q$, the disequilibrium is found to decrease with the increasing 
quantum number of the excited state. The amount of decrements increases near the
minima of disequilibrium versus $q$ plot.

According to Fig.~\ref{fig:qHO}c, with reference to the ground state,
for a given excited state, complexity of the q-oscillator has a very 
interesting and very complicated behavior. For all excited states complexity decrease 
depending on the increment in $q$, however, 
it starts to increase from above a critical minimum $q$ value.
This complication of the complexity is caused by the  
confliction between disequilibrium and information entropy. We can conclude that system 
has a critical transition in complexity.
Around the value of $q$ in the neighbourhood of 0.9 where 
information entropy exhibits a maximum and the disequilibrium exhibits a minimum, the
statistical complexity attains an approximately similar value for all excited states. 
Below this bunching point of statistical complexity for the excited states, 
the  $C_{LMC}$ vs $q$ curves are found 
to cross over with the inversion of relative ordering of the quantum number n of the 
excited states. 
Beyond the bunching region, as $q$ decreases, for each excited state the statistical 
complexity goes through a minimum at a certain $q$.
The depth of the minimum value is found
to decrease with decreasing quantum number of the excited state.
With further decrease in $q$ beyond the minimum 
value statistical complexity, the $C_{LMC}$ values are found to increase, finally, 
culminating at a common value at the lowest $q$.

Another interesting result that appears in q-oscillator is the following,
for low $n$ values, entropy takes low values, however, disequilibrium is also large.
Then complexity also takes large values for low $n$ values.

Finally, in order to comprehend the numerical changes, Shannon information entropy, 
disequilibrium and complexity values
obtained by using Eqs.~\ref{eq:C}-\ref{eq:D} are given in
Table \ref{tab:qHO_SDC}. 

\begin{table}[h!]
\centering
\caption{Energy level and q-deformation 
dependence of Shanon information entropy, disequilibrium and complexity 
for q-deformed harmonic oscillator.}
\label{tab:qHO_SDC}
\begin{tabular}{@{}llllll@{}}
\hline
$n$ & $q$ & $S$ & $D$ & $C=e^SD$ \\ 
\hline
0  & 0.001 & 1.07236 & 0.39894 & 1.16582 \\
& 0.4   & 1.07236 & 0.39894 & 1.16582 \\
& 1     & 1.07236 & 0.39894 & 1.16582 \\
5  & 0.001 & 1.07829 & 0.39232 & 1.15329 \\
& 0.4   & 1.59322 & 0.21132 & 1.03962 \\
& 1     & 1.76806 & 0.19666 & 1.15235 \\
10 & 0.001 & 1.07829 & 0.39232 & 1.15329 \\
& 0.4   & 1.61096 & 0.20561 & 1.02962 \\
& 1     & 2.01018 & 0.15668 & 1.16957 \\
\hline
\end{tabular}
\end{table}	
	
\subsection{q-Deformed Morse potential}\label{}

After quantifying the complexity of q-oscillator, here we consider, as an 
second example, q-deformed Morse
potential \cite{Cooper1995,Ikhdair2009,Dobrogowska2013,Vinh2015}.
The general form of the q-deformed Morse potential is expressed in the 
Ref.\,\cite{Znojil1999}  as the following,
\begin{equation}
	V_{MP}(x)=V_{1}e^{-2\alpha x}-V_{2}e^{-\alpha x},\; V_{1}=D_e,\;V_{2}=2qD_e
\end{equation}
where $\alpha =ar_{e},$ $x=(r-r_{e})/r_{e},$
$V_1$ and $V_2$ are the repulsive and attractive terms, respectively.
$r_{e}$ is the equilibrium position of the nuclei,
$a$ is a constant related with the range of potential and
$D_{e}$ is a measure of the depth of potential well at equilibrium distance.
This form of the q-deformed Morse potential is the special case of the well-known
diatomic Morse potential.  

Wave function of a particle for q-deformed Morse potential can be obtained 
after several steps as in Ref.\,\cite{Ikhdair2009}
	\begin{equation} \label{psi-mp}
	\Psi_n(x)=N_n\exp \left\{ -\alpha \left( \lambda q-n-\frac{1}{2}\right)
	x-\lambda e^{-\alpha x}\right\} L_{n}^{2\left( \lambda q-n-\frac{1}{2}%
		\right) }\left( 2\lambda e^{-\alpha x}\right)
	\end{equation}%
	where $N_n$ is the normalizing factor.
	$L_n^\alpha(x)$ are called generalized Laguerre polynomials or associated 
	Laguerre polynomials. In this work, we numerically computed the 
	normalization constant $N_{n}$ by using Mathematica. 
On the other hand, the energy eigenvalues of q-Morse potential are given as
\begin{equation} \label{m-energy}
	E_{n}=-\alpha ^{2}E_{0}\left[ \lambda q-n-\frac{1}{2}\right] ^{2},\text{ }%
	n=0,1,2,\cdots ,n_{\max }
\end{equation}
where $\lambda$ and $E_0$ are given as the following form,
\begin{equation}
	\lambda=\left(\frac{D_e}{\alpha ^{2}E_{0}}\right)^{1/2},\;
	E_{0}=\frac{\hbar ^{2}}{2\mu r_{e}^{2}}
\end{equation}
here $\mu=m_1 m_2/(m_1+m_2)$ is the reduced mass of the
diatomic molecule. As it can be seen from Eq.\,\ref{m-energy}, the form of 
the energy is clearly different from that of the classical 
Morse potential one. For example; energy levels corresponding to 
q-deformed Morse potential are upper bounded by $q$ and maximum level number is 
restricted by the inequality $n_{max}\leq (\lambda q -1/2)$.

Here, we also investigate $q$ dependence of the complexity, disequilibrium and 
information entropy for q-Morse potential. However, in the present case, we compute and 
discuss these quantities by using well known experimental parameters of HCL 
and H$_2$ molecules, which are given in Table~\ref{tab:molecules}. Here we note that, in 
numerical procedure, we defined $q=0.35$ as a lower bound. Under this lower bound of $q$, 
numerical error appears in computation.
Therefore, we study $q$ dependence between 0.35 and 1.0.
However, on the other hand, the lower bound of the $q$ leads to an upper bound which 
covers the number of possible excited states. For example, 
by using these values, in our work for $q=0.35$,
$n_{max}$ value is found as 8 and 5 for HCl and $\rm H_2$ molecules, respectively.

\begin{table}[!ht]
\centering
\caption{Molecular specific parameters for HCl and $\rm H_2$.}
\label{tab:molecules}
\begin{tabular}{lccc}
	\hline
	Parameter & HCl \cite{Qiang2007} & $H_2$ \cite{Flavio2015} & Unit \\
	\hline
	a & 1.868 & 1.944 & $\AA^{-1}$ \\
	$r_e$ & 1.275 & 0.742 & $\AA$ \\
	$D_e$ & 37255 & 38266 & $\rm cm^{-1}$ \\
	$\mu$ & 0.980 & 0.504 & amu \\
	\hline
\end{tabular}
\end{table}
	
\begin{figure*}[h!]
		\centering
\subfloat[$n=1$]
{\includegraphics*[width=0.4\textwidth]{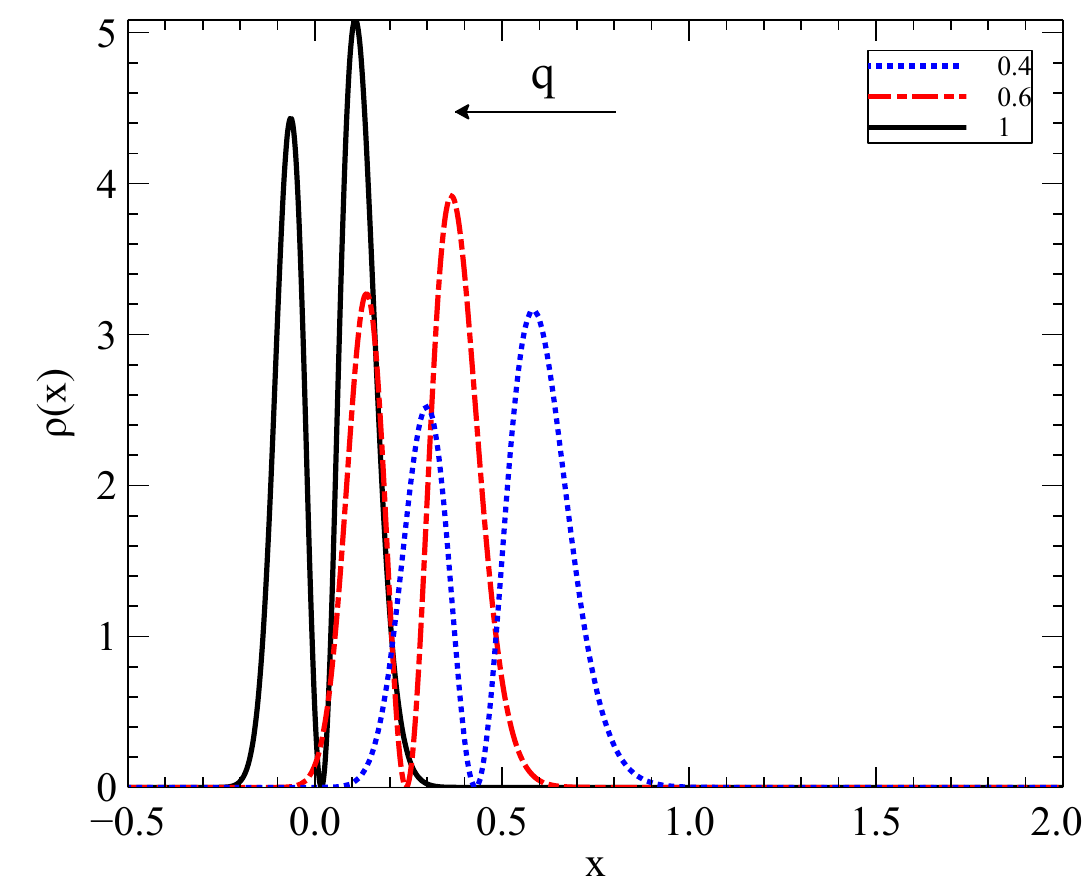}}
\subfloat[$n=4$]
{\includegraphics*[width=0.4\textwidth]{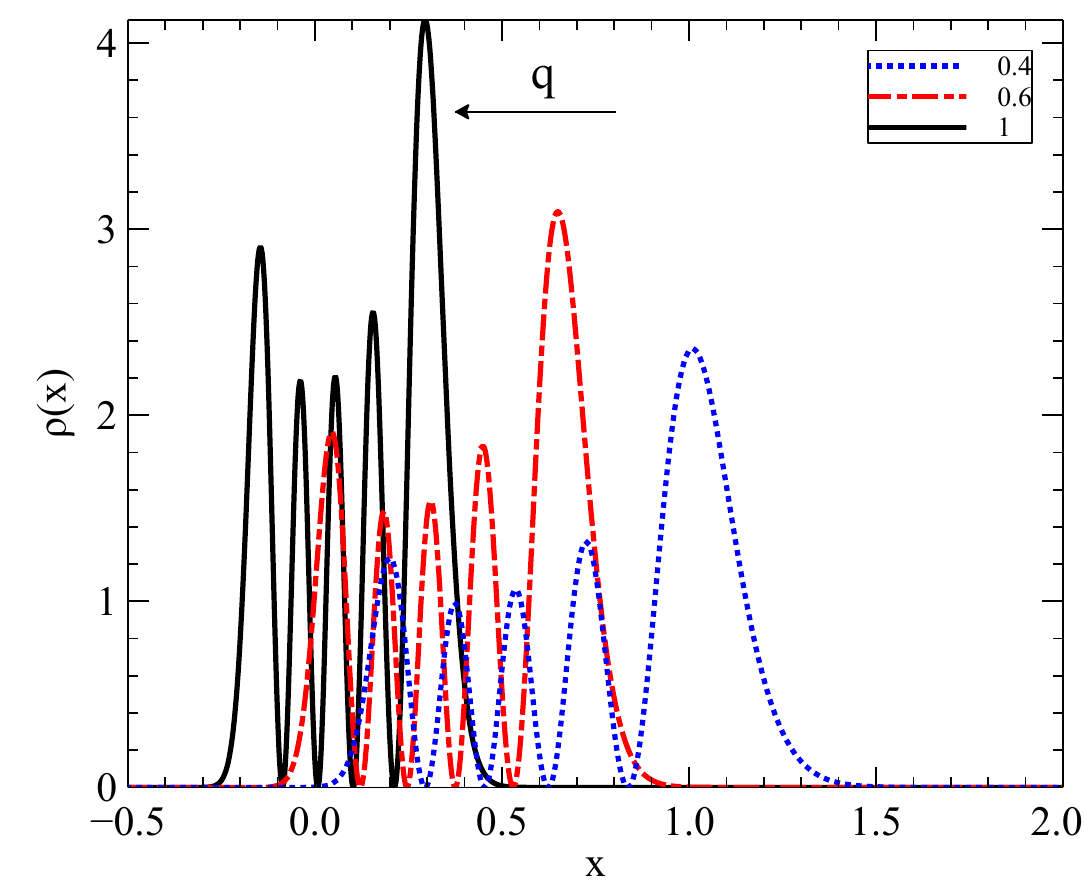}}
\caption{$q$ dependence of the probability distribution under q-deformed 
Morse potential for levels $n=1$ and $n=4$ of HCl molecule.
The arrow indicates increment direction of $q$.}
\label{fig:qMP_HCl_rho}
\end{figure*}

\begin{figure*}[h!]
		\centering
\subfloat[$n=1$]
{\includegraphics*[width=0.4\textwidth]{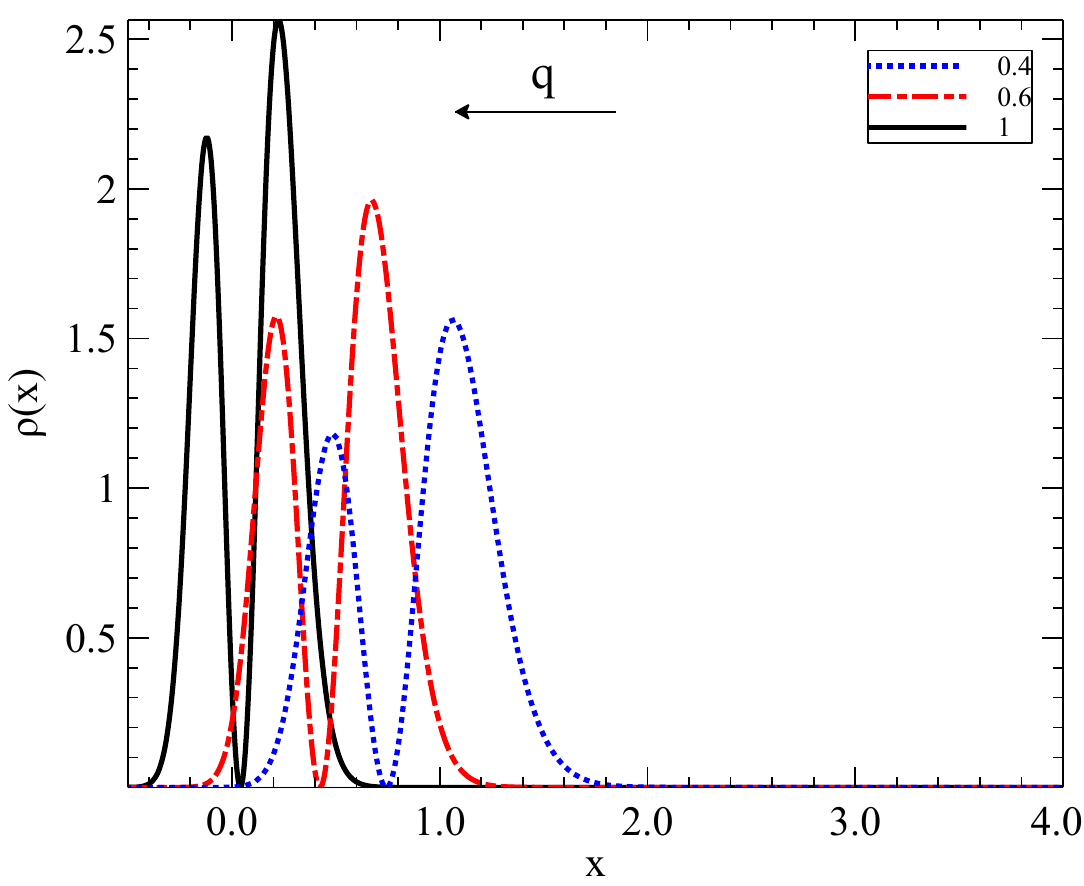}}
\subfloat[$n=4$]
{\includegraphics*[width=0.4\textwidth]{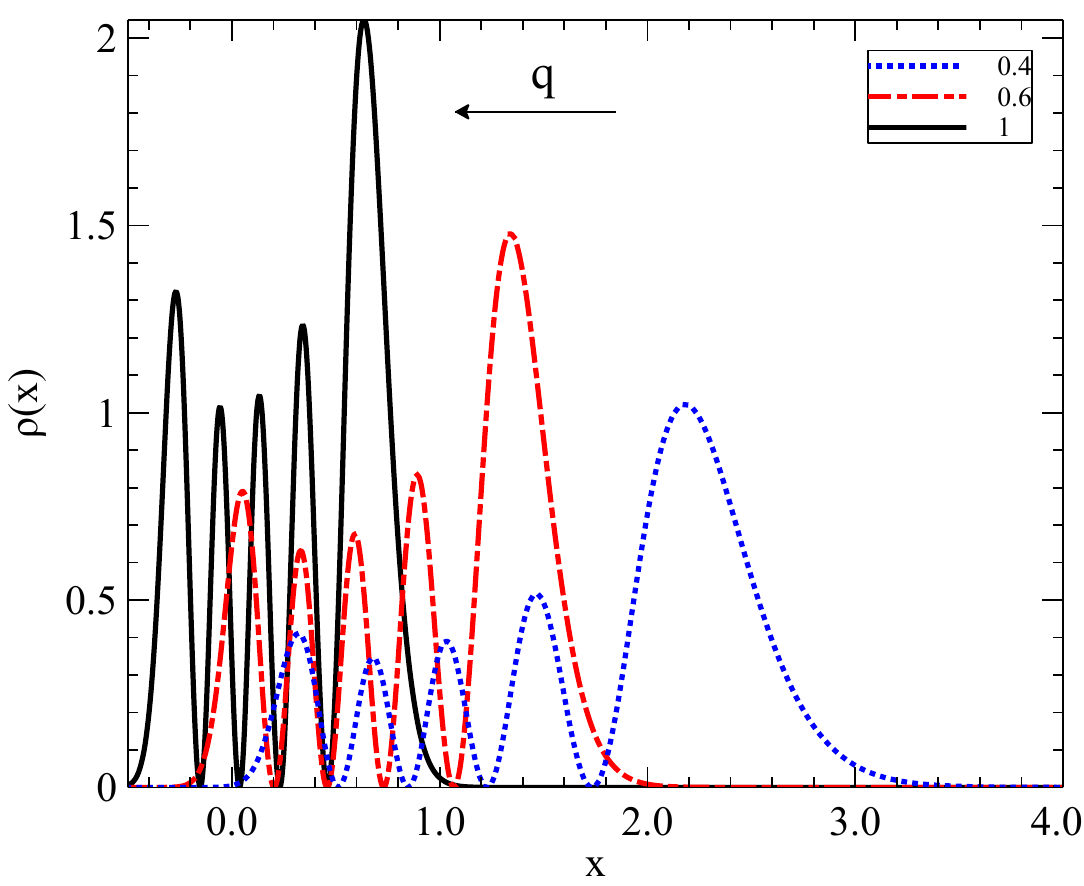}}
\caption{$q$ dependence of the probability distribution in q-deformed 
harmonic oscillator for levels $n=1$ and $n=4$ of $\rm H_2$ molecule.
The arrow indicates increment direction of $q$.}
\label{fig:qMP_H2_rho}
\end{figure*}

The spatial probability distribution change with the q-deformation
parameter is presented for HCl and $\rm H_2$ molecules in 
Figs.~\ref{fig:qMP_HCl_rho} and \ref{fig:qMP_H2_rho}, respectively.
As it can be seen from figures that probability distribution of Morse 
potential has an asymmetric shape.
For $q=1$ peaks take large values however, when 
the $q$ parameter decreases $\rho(x)$ becomes much more spreading over position space. 
This change can be more easily seen in excited states such as shown in 
Figs.~\ref{fig:qMP_HCl_rho}b and \ref{fig:qMP_H2_rho}b for state $n=4$.

\begin{figure*}[h!]%
	\subfloat[]
	{\includegraphics*[width=0.32\textwidth]{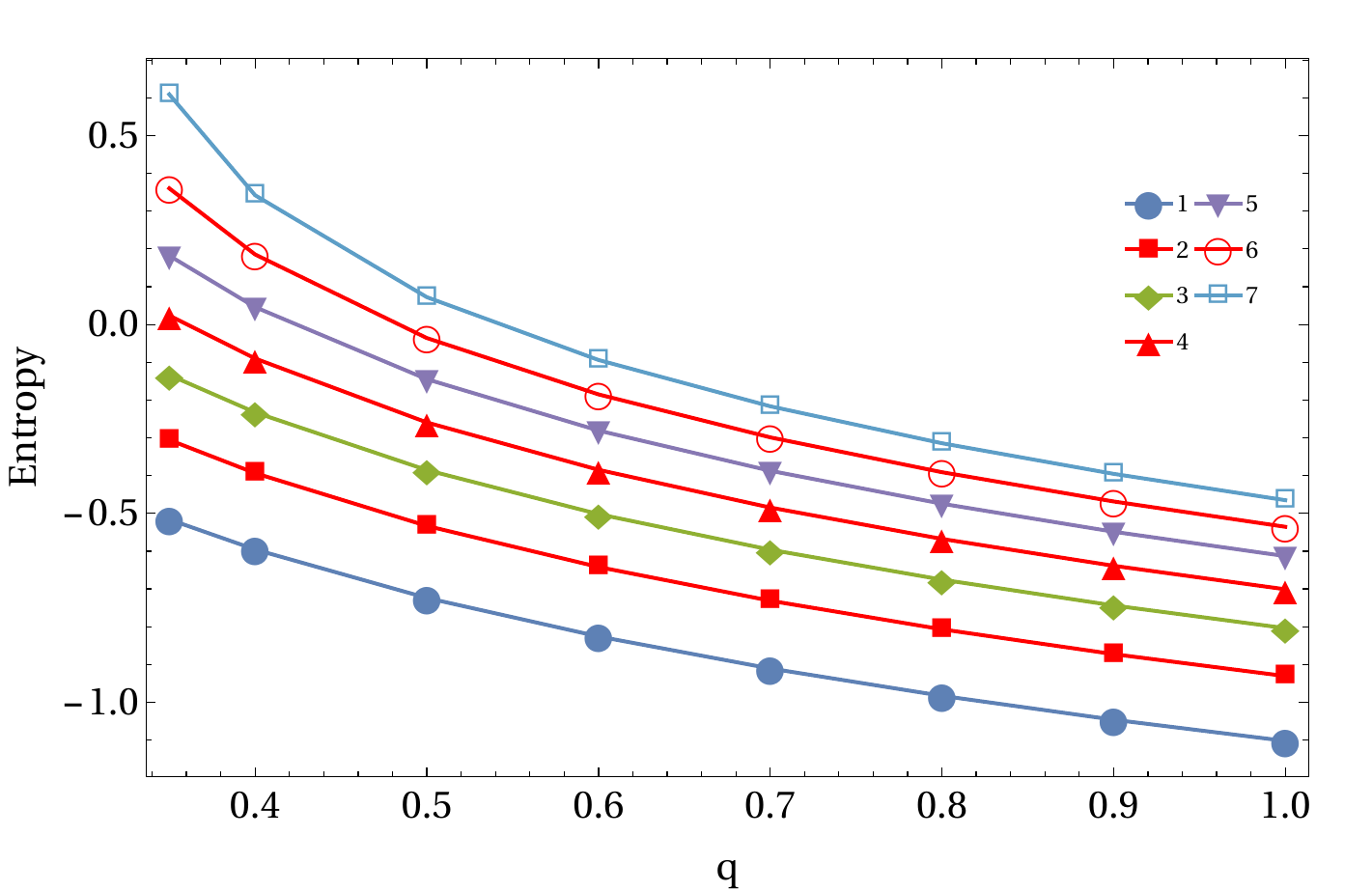}}
	\subfloat[]
	{\includegraphics*[width=0.32\textwidth]{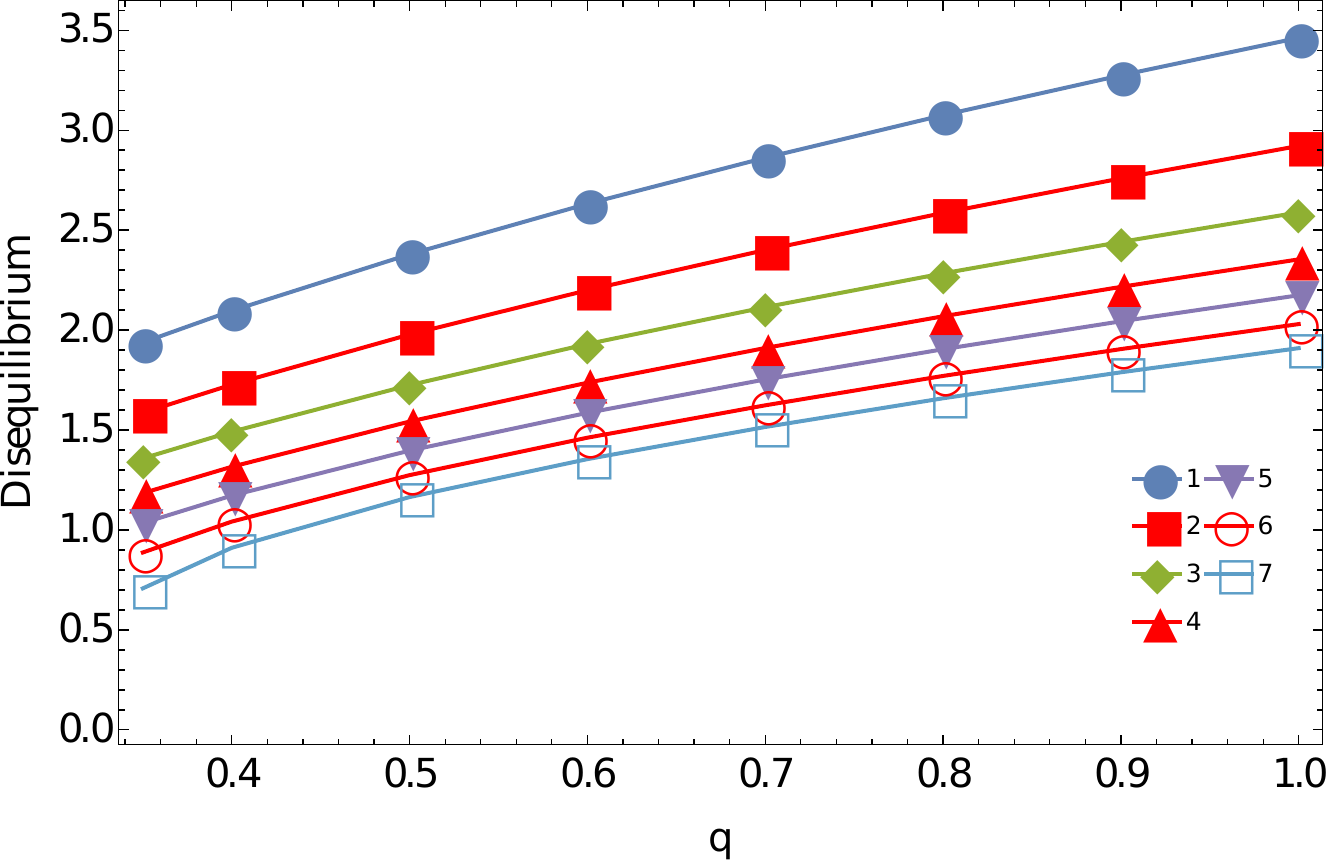}}
	\subfloat[]
	{\label{fig:figs4c}%
		\includegraphics*[width=0.32\textwidth]{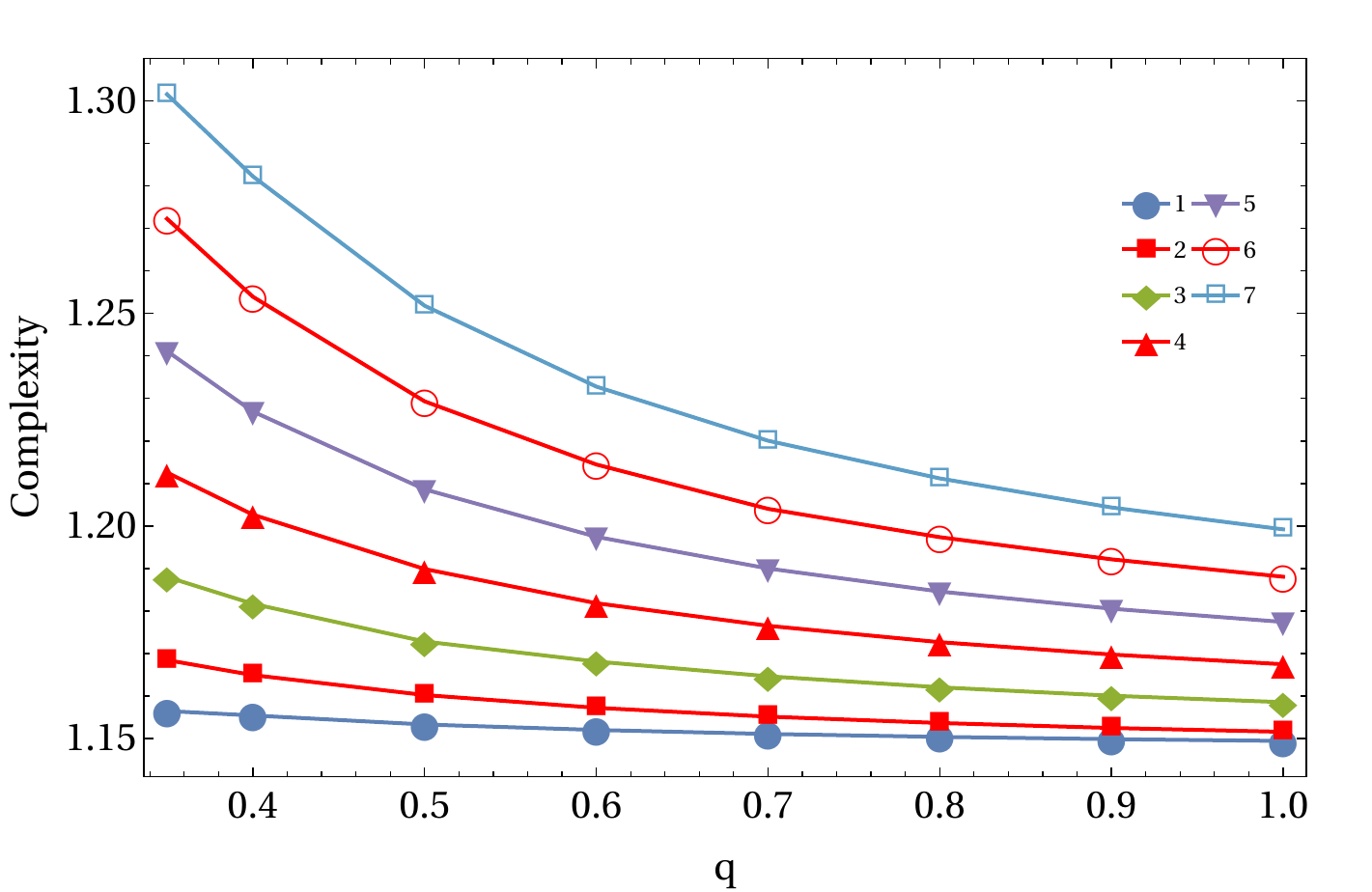}}%
	\caption{Shanon information entropy (a), disequilibrium (b) and complexity 
		(c) for HCl molecule at energy levels n in [1,7].}
	\label{fig:qMP_HCl}
\end{figure*}
\begin{figure*}[h!]%
	\subfloat[]
	{\includegraphics*[width=0.32\textwidth]{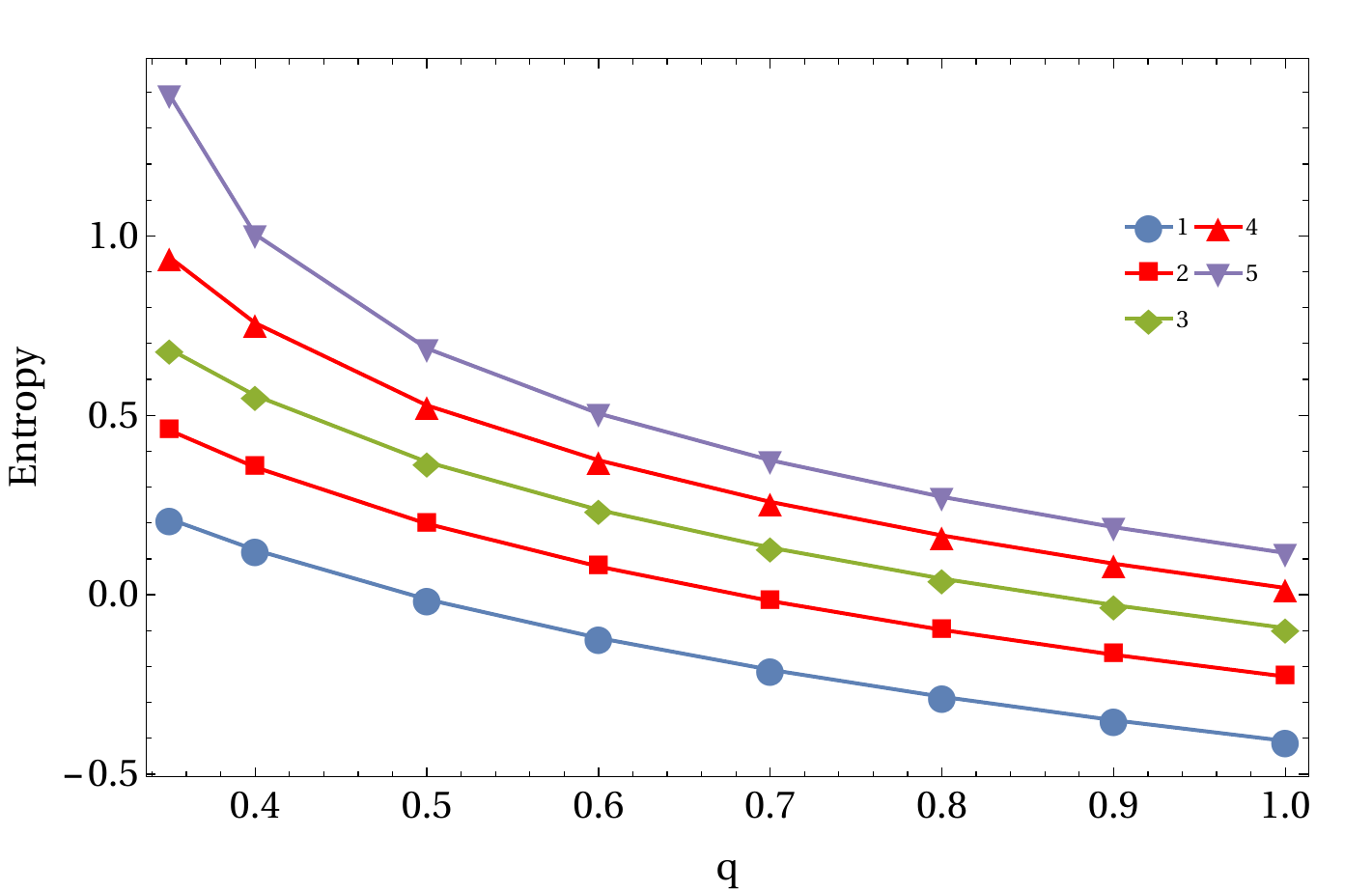}}
	\subfloat[]
	{\includegraphics*[width=0.32\textwidth]{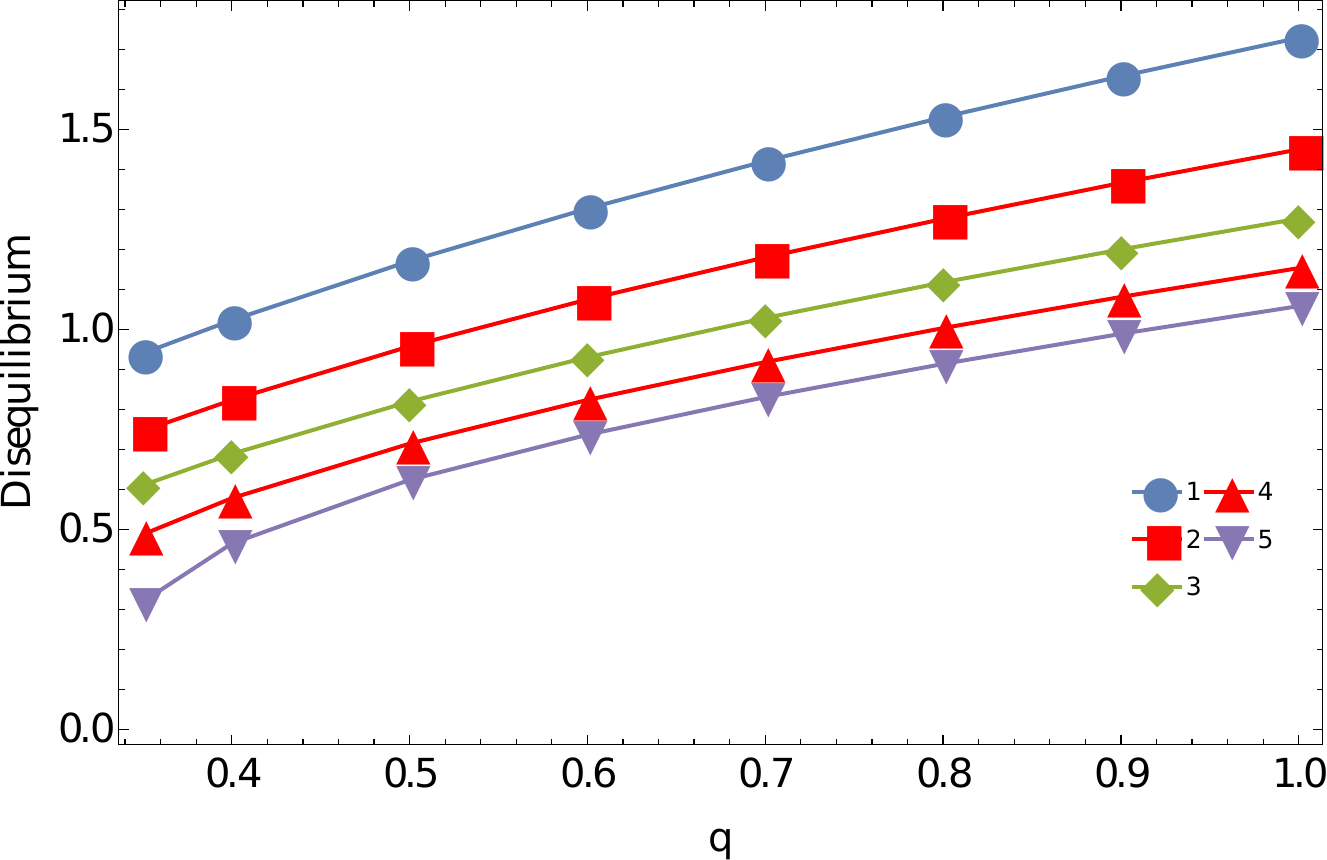}}
	\subfloat[]
	{\includegraphics*[width=0.32\textwidth]{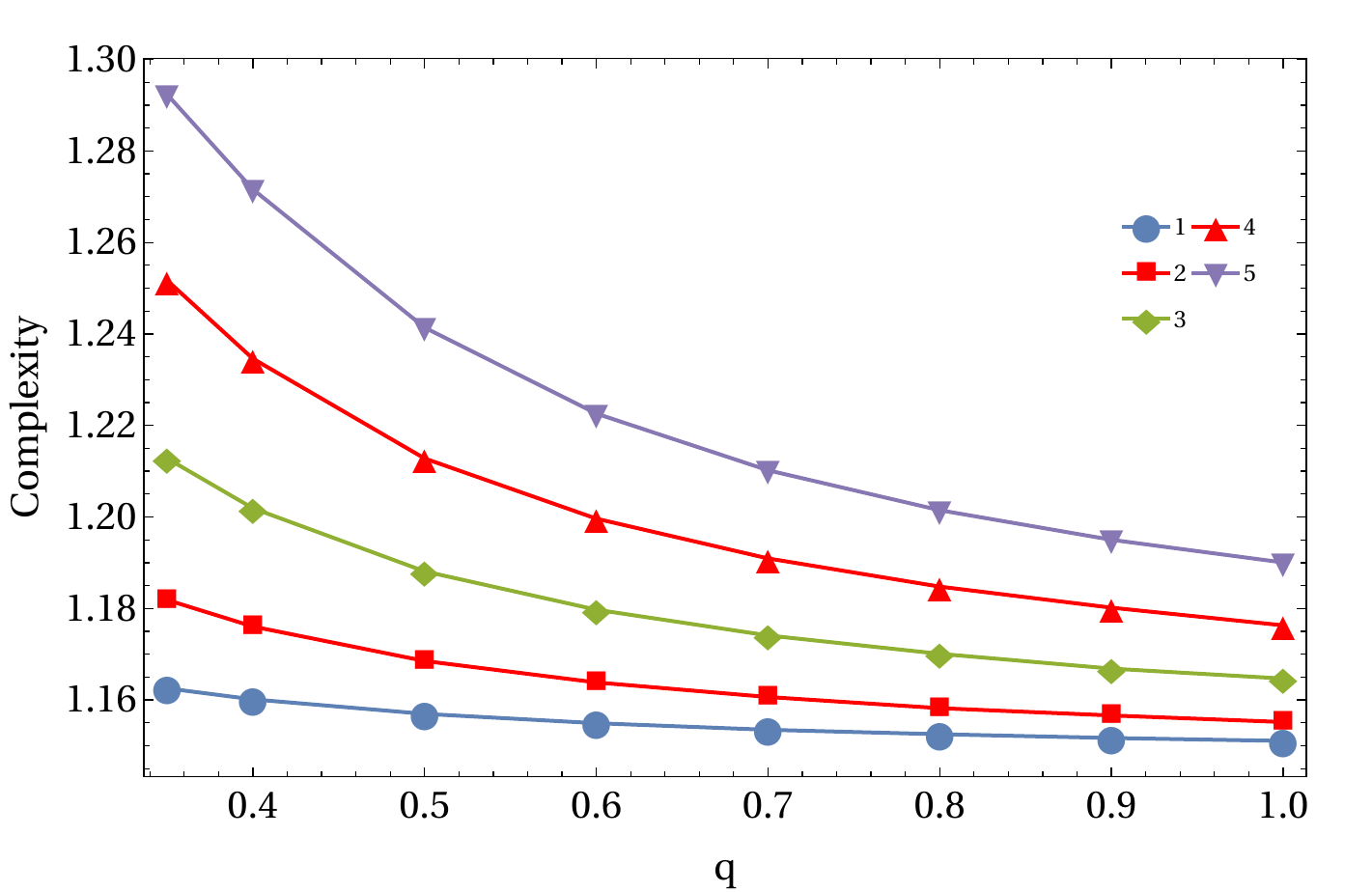}}%
	\caption{Shanon information entropy (a), disequilibrium (b) and complexity 
		(c) for $\rm H_2$ molecule at energy levels n in [1,5].}
	\label{fig:qMP_H2}
\end{figure*}
Here we compute Shanon information entropy, 
disequilibrium and complexity by using solutions of q-deformed Morse potential for HCl 
and $\rm H_2$ diatomic molecules. By using parameters in Table\,\ref{tab:molecules},
we compute  all quantities.
The numerical results are given in Figs.~\ref{fig:qMP_HCl}
and \ref{fig:qMP_H2} for several excited states for HCl and $\rm H_2$ molecules,
respectively.
As it can be seen from figures that there is no any complicated behavior in entropy, 
disequilibrium and complexity as in q-harmonic oscillator. 
For both molecules, entropy $S$ and complexity $C$ smoothly decreases with $q$, however, 
disequilibrium $D$ smoothly increases with $q$. 

Now we can compare some result of the q-deformed Morse potential with the 
q-oscillator. When comparing results we can see that for low $n$ values entropy takes low 
values, however disequilibrium is also large.
These results are compatible with the results of 
q-oscillator. However, the complexity of q-Morse takes low values for low n values unlike 
q-oscillator.

\section{Concluding remarks}\label{sec:conclusions}

When q-deformation applied to a quantum harmonic oscillator,
it is found that probability distribution becomes more classical
rather than quantum mechanical as $q$ approaches to zero.
In q-deformed harmonic oscillator, as $q$ decreases
$\rho(x)$ becomes more localized.
$q$ deformation enforces q-oscillator to behave like a classical
oscillator without spreading the probability distribution but preserving
the discreteness of energy.
In other words, in the limit case in which deformation increases,
$q$ approaches to 0, probability distribution
of a single particle collapses in to a Gaussian form.
$q$-deformation phenomena reduces the Heisenberg uncertainty
$\Delta x \Delta p$ product  for the
sake of gradual lose of quantumness.
Entropy, disequilibrium and complexity values are independent of $q$ for the ground state
of the q-oscillator. However, q-deformation causes a $q$
dependent change in entropy, disequilibrium and complexity 
calculated for excited states.
q-deformation affects the oscillator differently 
depending on its energy level $n$. For instance, minima
of the complexity shift to higher $q$ values as the quantum
number $n$ increases. q-oscillator takes the same complexity value,
which is calculated for position space,
independent of its energy when $q$ is set to 0.9.

Complexity measure of probability distribution
of an electron belonging to a q-deformed
diatomic Morse potential decreases with increasing $q$.
Meanwhile, when $q$ is increased, entropy decreases and disequilibrium
increases. When q-dependence of the probability distributions of position space
for q-harmonic and q-Morse potentials are compared
it is seen that 
q-deformation (as $q$ approaches zero) causes localization in q-oscillator
by squeezing $\rho(x)$, whereas it causes a spreading of $\rho(x)$ under 
q-Morse potential.

Complexity measure of probability distribution
of an electron belonging to a q-deformed
diatomic Morse potential decreases with increasing $q$.
Meanwhile, when $q$ is increased, entropy decreases and disequilibrium
increases. When q-dependence of the probability distributions of position space
for q-harmonic and q-Morse potentials are compared
it is seen that 
q-deformation (as $q$ approaches zero) causes localization in q-oscillator
by squeezing $\rho(x)$, whereas it causes a spreading of $\rho(x)$ under 
q-Morse potential.

\section{Acknowledgments}\label{sec:acknowledgments}
KDS is grateful to The Scientific and Technological Research Council of Turkey 
(T\"{U}B\.{I}TAK) for a visiting scientist award under its 2221-program, grant number 
1059B211601794 and acknowledges with thanks the support received under the Emeritus 
Scientist scheme , C.S.I.R. New Delhi.

%\begin{acknowledgments}
%\end{acknowledgments}

% \nocite{*}

\bibliography{references}% Produces the bibliography via BibTeX.

\end{document}